\documentclass[12pt,a4paper]{article}

\usepackage{amsmath}
\usepackage{graphicx}
\usepackage{times}
\usepackage{cite}
\usepackage{ulem}
\begin{document}
\begin{titlepage}
\title{Separating energy scales in hadron scattering}
\author{ S.M. Troshin, N.E. Tyurin\\[1ex]
\small  \it NRC ``Kurchatov Institute''--IHEP\\
\small  \it Protvino, 142281, Russian Federation,\\
\small Sergey.Troshin@ihep.ru
}
\normalsize
\date{}
\maketitle

\begin{abstract}
	We discuss energy scales in soft hadron scattering and  their separation: the one  relevant to the region where the total cross--sections  increase starts and another one related to the asymptotic region. The latter  is most sensitive to the unitarity effects and the former one  --- to the gross features of absorption of the initial wave due to multiparticle production.  Transition  from  the shadow to  reflective scattering sets a relation between the two energy scales.
\end{abstract}
\end{titlepage}
\setcounter{page}{2}
\section*{Introduction}

It is a notorious statement that the study of soft hadron interactions in absence of a reliable theoretical approaches has to address   the general principles such as unitarity and analyticity    to provide a consistent description.  The existing models should take these principles into account but cannot pretend to be unique theoretical schemes  even being the QCD-inspired   and despite of their often widespread use. 

Approach to the asymptotic regime depends on a specific model being used and thus there is an energy scale where the asymptotics   becomes effective. Solution of this case in a model--independent way is not trivial,  and the discussion  can be found in \cite{strum} and \cite{wall}.

On the other hand, there are the experimental indications on transition to the reflective scattering mode \cite{trans} with the energy increase. The energy value where this transition starts can be hopefully used as a reference point for  separation between  the asymptotic region with the unitarity effects domination and the region of lower energies   where hadron interaction dynamics responsible for the total cross--section growth prevails. 
Indeed, this energy range determines a noticeable and important scale in hadron interactions.
Namely,
the  measurements at the LHC energy of $\sqrt{s}=13$ TeV  have indicated \cite{tamas} that  the  interaction region attributed to inelastic processes is transforming from a black disk   to a black ring.  The results of the two independent analyses \cite{tamas} and \cite{koval} are consistent, and the both confirm exceeding
the black disc limit.
Transition to   black ring picture  has been discussed in \cite{trans} and references therein.
Such a picture naturally emerges from the reflective scattering mode (antishadowing) \cite{plb93} being its higher energy consequence. 

To clarify this statement  we turn to
 the unitarity relation for the scattering amplitude in the impact parameter representation  
$f(s,b)$  written in the form:
\begin{equation}\label{unit}
\mbox{Im}f(s,b)[1-\mbox{Im}f(s,b)]=h_{inel}(s,b)+[\mbox{Re}f(s,b)]^2
\end{equation}
where $h_{inel}(s,b)$ is the inelastic overlap function, which takes nonnegative values.  Neglecting the real part of the scattering amplitude \cite{real} and making the respective replacement $f\to if$,  Eq. (\ref{unit}) can be written as follows:
\begin{equation}\label{unit1}
f(s,b)[1-f(s,b)]=h_{inel}(s,b).
\end{equation}
The LHC measurements of the elastic scattering indicated  that the transition matrix element $S(s,b)$ is close to zero in the interval $0<b\leq 0.4$  fm \cite{tamas}. Then, following the relation \cite{prd13}:
\begin{equation}\label{unit2}
\frac{ \partial h_{inel}(s,b)}{\partial s }=S(s,b) \frac{\partial f(s,b)}{\partial s}
\end{equation}
we conclude that
the energy evolution of $h_{inel}(s,b)$ is suppressed in comparison with  evolution of the amplitude $f(s,b)$ in the range where $S(s,b)\simeq 0$. The scattering amplitude $f(s,b)$ is more sensitive to switch-over from the shadow to reflective scattering mode and numerical analysis \cite{alkin} indicated the  transition energy $\sqrt{s_r}$ might be  below of $7$ TeV, and indeed the black ring formation  occurs at higher energies \cite{tamas}.  Of course, those estimates  are to be taken with care 
due to a numerical smallness of the observed effect and  the different options used for  the scattering amplitude reconstruction from the  experimental data.

\section{Unitarity effects and asymptotics}
It seems reasonable to assume   $\sqrt{s_r}=5$ TeV as an energy value where $f(s,b=0)$ exceeds the value of $1/2$ and thus enters the reflective scattering mode.  The two regions are depicted at Fig. 1.
 \begin{figure}[hbt]
	\vspace{-0.5
	cm}
	\hspace{0.9cm}
	%	\begin{center}
	\resizebox{12cm}{!}{\includegraphics{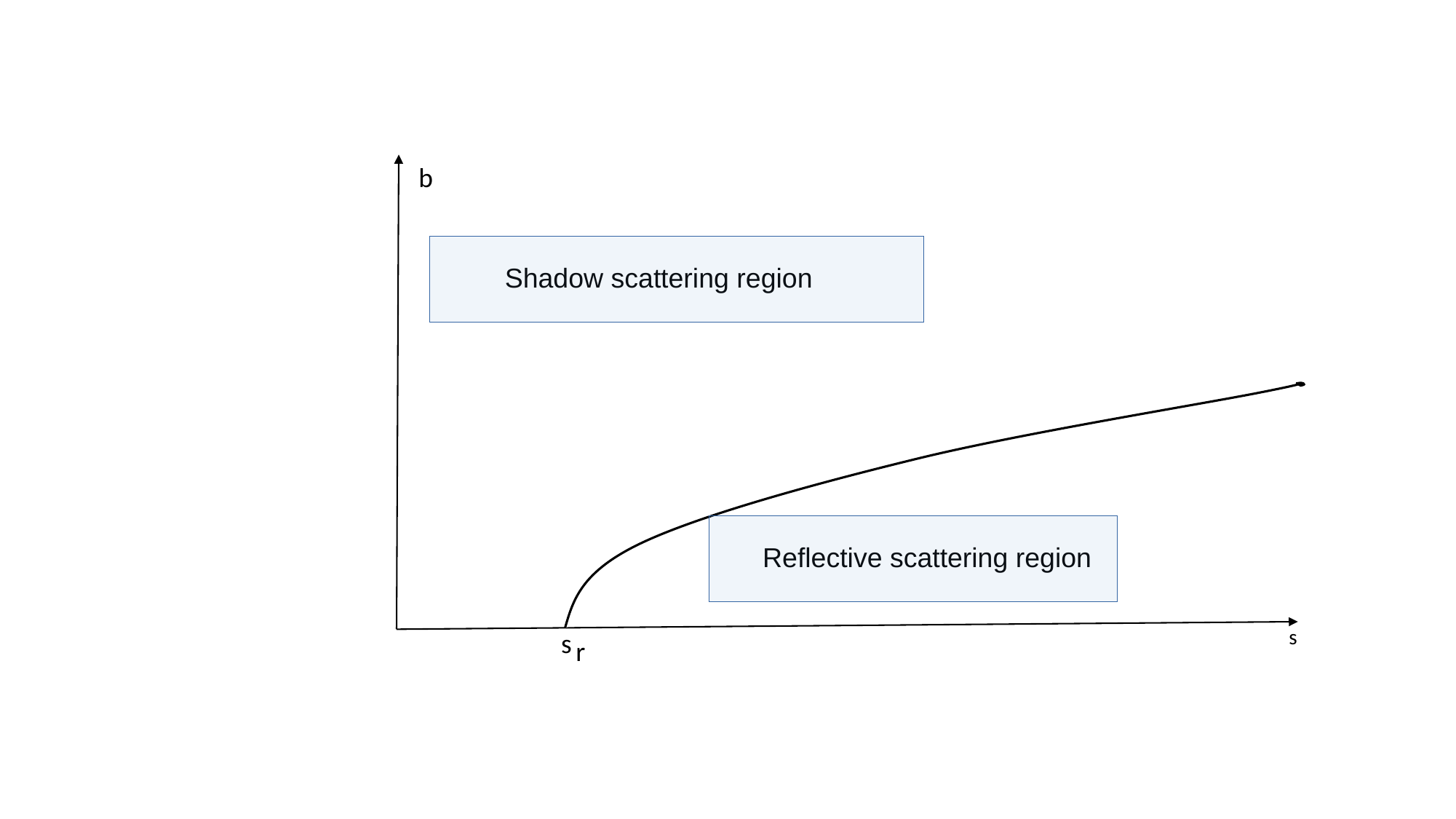}}		
	%\includegraphics{dsdb2n.eps}
	%	\end{center}
	\vspace{-1.2cm}
	\caption{Schematic representation of the shadow and reflective scattering areas.}	
\end{figure}
Accepting  $\sqrt{s_r}=5$ TeV as an energy  value where transition to the asymptotics  begins, one can assume that the values of $\sqrt{s}\gg\sqrt{s_r}$ determine the region,  where the unitarity effects dominate and are decisive for hadron interactions dynamics. Thus, it would be natural to
to estimate by the value of $10 \sqrt{s_r}\simeq 50$ TeV the energy range where one would expect the  following  asymptotical dependencies of the global observables \cite{plb93}:
\begin{equation}\label{ln2}
\sigma_{tot}(s)\sim \sigma_{el}(s)\sim B(s) \sim \ln^2(s),
\end{equation}
\begin{equation}\label{ln1}
\sigma_{inel}(s)\sim \ln(s),\;\;\;
\rho(s)\sim \ln^{-1}(s).
\end{equation} 
and  the effective intensity of interactions $Y(s)\equiv\sigma_{tot}(s)/16\pi B(s)$ increases and approaches its limiting value of unity \cite{int}. Indeed, experimentally, observed  ratio  of $B(s)/\sigma_{tot}(s)$ decreases at available energies \cite{shan}. Note, that $Y(s)<1$.

The asymptotic   dependencies, Eqs. (\ref{ln2}) and (\ref{ln1}),  are associated with the reflective scattering mode and are the consequence  of self-damping of the inelastic channels in this mode \cite{baker}.  The self--damping of  inelastic channels    leads to the limit 
\begin{equation}\label{sdamp}
\lim _{s\to\infty}\sigma_n(s,b)=0,
\end{equation}
for the partial cross-section of $n$ produced particles and is valid for any fixed value of the impact parameter \cite{mdist}. It is to be noted that the available experimental data, for example, the increase of the ratio $\sigma_{el}(s)/\sigma_{tot}$ do not conflict with the Eqs. (\ref{ln2}) and (\ref{ln1}). 

It should also be noted that the value of $S(s,b)$ is negative under the reflective scattering mode and it leads to additives of opposite signs to the scattering amplitude  $f$ and the inelastic overlap function $h_{inel}$  under energy evolution in this mode at fixed impact parameter. The equality
\begin{equation}\label{unit3}
 h_{inel}(s,b_1)=h_{inel}(s,b_2)
\end{equation}
which takes place at two different values of the impact parameter $b_1<b_m<b_2$ ,where  
$h_{inel}(s,b_m)=1/4$ means that $S(s,b_1)=-S(s,b_2)$. The latter equality can be interpreted as an appearance of  an analog of Berry phase \cite{berry}  in the relativistic case, when $h_{inel}$ is considered as the  parameter for  two--particle  quantum system description (see also \cite{refb}). We would like to emphasize   that the underlying mechanism of this phase emergence here is related to  transition into the domain of strong coupling interactions  and will be discussed in Section 3.
\section{Growing role of inelastic channels with the energy increase}
The unitarity effects with  self-damping of inelastic channels  as well as the collective effects are significant at very high energies.  The elastic scattering amplitude at the energies $\sqrt{s}<\sqrt{s_r}$ has the shadow nature, i.e. its magnitude   straightforwardly reflects an increasing role of the newly opened inelastic channels contributions\footnote{The only shadow scattering is possible at the energies  $s<s_r$, see Fig.1.}. 

The  model  with independent emission of the secondary particles applied  for  calculation of the inelastic overlap function and  then the elastic amplitude extraction with a restriction of the shadow scattering mode \cite{vanh}  can be used for illustration of the above point.  Note, the shadowing in the model  could be interpreted as a consequence of an absence of  interactions under production of secondaries or their clusters in the limited ranges of the longitudinal momenta (``uncorrelated jets''). 

The respective transition between the two modes becomes particularly transparent when one represents the scattering amplitude (pure imaginary) in the impact parameter representation  as 
\begin{equation}\label{slin0}
f=h_{el}+h_{inel}=f^2+f(1-f)
\end{equation} 
and notes that crossing the value $f=1/2$ changes the prevailing contributing role of $h_{inel}(s,b)$. As it was noted, the elastic scattering in the region of $f<1/2$ has a shadow nature and the amplitude  reflects the dominating role of  inelastic channels contribution to Eq. (\ref{slin0}). 

Interactions of various hadrons  ($pp$, $\bar pp$, $\pi^{\pm}p$, $K^{\pm}p$) have been experimentally studied at the energies $\sqrt{s}< 500$ GeV. Increase of the  total cross--sections in this limited region of energy variation is well described by the universal dependence  
\begin{equation}\label{slin}
\sigma_{tot}(s)=a+c\sqrt{s}
\end{equation} 
with  reaction-dependent constants $a$ and $c$ \cite{nadol}. 
It is not surprising that the total cross--sections increase linearly with $\sqrt{s}$.  
Eq. (\ref{slin})  reflects an increasing number of the inelastic channels with the energy: 
\begin{equation}
\sigma_{tot}(s)=\sum_n^{n_{max}} \sigma_n(s),\,\, n_{max}\sim {\sqrt{s}/m_{\pi}}.
\end{equation}
The dependence, Eq. (\ref{slin}), allowed for a finite  range of energies can be associated with a lack of correlations under emission of the secondaries \cite{vanh}. It is interesting to note that
fluctuations of multiplicity $\delta n$ at these energies are  mainly due to fluctuations of the collision impact parameter value $\delta_b n$:
\begin{equation}\label{delt}
\delta n \equiv n-\langle n \rangle=\delta_b n+\delta_q n.
\end{equation}
Here $\delta_q n$ stands for the  quantum fluctuation of multiplicity at  fixed   impact parameter value. 

Both the unitarity and presence of  a mass gap (short  interaction range) in strong interactions transform $\sqrt{s}$     dependence, Eq. {\ref{slin}}, into $\ln^2 s$ at $s\to \infty$. This  slow down  is accompanied by appearance of the collective effects  such as ridge\footnote{Appearance of ridge in small systems at high multiplicities is expected to correlate with transition to the reflective scattering mode \cite{mdistr}.} and the anisotropic flows  \cite{wei}, and quantum fluctuations of multiplicity become a dominating  contribution with the energy growth \cite{mdist}.

It should be noted that  similar to Eq. (\ref{slin}) dependence takes place for the total cross-section of $\gamma^* p$--scattering  extracted from  the low-x behavior of the structure function $F_2(x,Q^2)$:
\begin{equation}\label{hera}
F_2(x,Q^2)=a(Q^2)+c(Q^2)/\sqrt{x}.
\end{equation}
Indeed, the total cross-section of $\gamma^* p$-interactions  also increases   as a linear function of $\sqrt{s}$ \cite{nadol}.

The dependence\footnote{$Q^2$ denotes the virtuality of $\gamma^*$.}, Eq. (\ref{hera}), is in agreement with the respective HERA data, which correspond to the c.m. energy range  of $50-300$ GeV for $\gamma^*p$ collisions.  
Among the various hadron interactions,  the most interesting one is associated with $pp$--interactions since the LHC provides collisions 0f the highest available  energy. It is this case is under discussion   and the numerical estimates are given for  $pp$--interactions.

The  experimental observation of the  strong increase  of the effective total cross--sections, Eq. (\ref{slin}),  at moderate energies correlates with the related increase of intensity of hadron interactions with the energy.

\section{U--matrix approach}
 Under model construction of the amplitude,  only partial restoration of the unitarity occurs often. The  examples are adoption of a black 
disc picture or the use of eikonal approximation.

The unitarity  can be naturally and fully accounted  in the form of the generalized reaction matrix  approach representing the elastic scattering amplitude by the equation \cite{ech}:
\begin{equation}\label{um}
F=U+iUDF
\end{equation}
expressed here in the operator form 
 with the $U$ being  an input under model construction\footnote{Definition of  operator $D$ can be found in \cite{ech}.}.
 Eq. (\ref{um}) is the relativistic generalization of the basic equation of the quantum theory of radiation damping \cite{hit}, which implies {\it depletion of the transient state} with the energy increase and provides solution of the unitarity for the  amplitude in impact parameter representation in the rational form 
\begin{equation}\label{f}
f(s,b)=u(s,b)/[1-iu(s,b)],
\end{equation}
where the complex  function $u$  is a subject for  modeling with the only constraint $\mbox{Im}u\geq 0$  \cite{prd}.  With this constraint, it is a direct solution to the  unitarity  for the scattering amplitude $f$, Eq. (\ref{unit}). Eq. (\ref{f})   provides one-to-one mapping of the functions $u$ and $f$. The function $\mbox{Im}u$ gets contributions from the inelastic channels only and is therefore a result of the inelastic processes.
Eq. (\ref{f}) covers {\it the full unitary circle}, i.e. it covers the  both discussed modes -- shadow and reflective ones \cite{map}.  
 We use the
 approximation for the scattering amplitude based on dominance of its imaginary part which means   vanishing of a real part   of the input function $U$ and the  replacement $u\to iu$.

The function $u$ is proportional to the integral over the respective scattering potential  \cite{gold}
 \begin{equation}\label{bg}
 	u\propto \int_{-\infty}^\infty V[\sqrt {(z^2+b^2)}]dz .
 \end{equation}
Its relation with the quasi-potential \cite{logt} as well as its possible form as a superposition of the Yukawa potentials  has been discussed  in  \cite{ech}. The function $u(s,b)$   can be treated as a generalized potential with  energy--dependent coupling in  high energy particle scattering (see also ref. \cite{pot}).

 Thus, the amplitude   varies  in the shadow region, $0<f<1/2$, when $u<1$ (weak coupling),  and in the reflective region, $1/2<f<1$ when $u>1$ and $S$ is negative, i.e. the function $S$ acquires phase $\pi$. So, the solution, Eq. (\ref{f}), includes the both weak and strong couplings and is suitable for consideration of the nonperturbative region of interactions.

 We would like to make a comment here on the eikonal approximation for the scattering amplitude\footnote{Evidently, eikonal approximation is not an equivalent to the exponential representation of  amplitude as a complex quantity.}. It implies an existence of  shadow scattering mode (weak coupling regime) only.  As it is  known, this high energy approximation is based on the assumption that  high energy particle undergoes a small deflection during scattering process due to a relative weakness of the interaction potential. This assumption excludes the reflective scattering mode having a nonperturbative origin. 
 
  Notion of  a potential must be generalized  for the case of   high energy relativistic particle scattering when  the potential should  become an energy -- dependent quantity.  The infinitely rising total cross-sections imply that the potential rises with the energy too. Thereby assumption on a small deflection of  high energy particle during the scattering cannot be  justified at high energies and so is the standard eikonal approximation itself (but, of course, not an exponential representation of the scattering amplitude). 
This  is correlated with the result of paper \cite{bart},  where it has been shown  that the eikonal approximation is valid in the limit of weak coupling only and it implies neglect  of the scattering particles with large transverse momenta. Extending  the eikonal approximation seems to be doubtful  and its use  e.g. for calculation of gap survival probability   can  result in  a misleading conclusion on the reflective scattering mode \cite{khoze}.
  
 The reflective scattering mode must be presented at high energies under description of the strong interactions dynamics.
Thus, presence of the two scattering modes --- shadow and reflective --- seems to be relevant issue for  a  theory of strong interactions which inevitably implies a nonperturbative mechanism.
 \section*{Conclusion}
Following the above considerations one can infer that the shadow and reflective scattering modes correspond to a weak and strong couplings of  hadron interactions theory, respectively, and appearance of the reflective scattering is a natural consequence of transition to  strong coupling  implied by the unitarity and  interaction dynamics.
 This conjecture finds its justification  in the $U$--matrix form of the amplitude unitarization, where the input function $u(s,b)$ grows up with the energy. This growth  reflects the key properties of hadron interactions. The ultimate asymptotic regime expressed by Eqs. (\ref{ln2}) and (\ref{ln1}) as well as the total cross--section increase described by the universal empirical dependence, Eq. (\ref{slin}),  are related to a single mechanism considering the scattering amplitude as solution of Eq. (\ref{um}) which naturally accounts for  presence of the two scattering modes with  different dynamical appearances of the inelastic channels contribution. Both of these modes are consistent with the unitarity.

 The amplitude being  solution of Eq. (\ref{um}) provides a smooth, continuous transition between the two  scattering modes. The rational form of unitarization (U--matrix)  can be treated as a  way  of  transition  to the strong coupling limit in hadron scattering.
The energy range  between these two modes separates 
the asymptotical region  and the  range of lower energies dominated by  dynamics that determines the total cross--section increase.  This energy gap sets a relation between the two respective energy  scales. 

It is to be noted that existence of the reflective scattering mode is in accord with  Chew and Frautchi postulate on the maximal strength of hadron interactions: `` a characteristic  in the strong interactions is   a capacity to ``saturate'' the unitarity conditions at high energies'' \cite{cfr}. 

The  reflective scattering mode  along with the shadow scattering provide the required consistency and completeness in hadron interactions at  high energies. {\it Ad hoc } exclusion of the reflective scattering mode would mean an unjustified assumption on neglect of the strong coupling. 
The interaction dynamics is determined by QCD  theory  with colored objects (quarks and gluons) confined inside
hadrons \cite{ji}. The confinement phenomenon is responsible for the  strength of soft hadron interactions and appearance of the collective effects  \cite{mdistr}. The relevant scale  $\Lambda_{QCD}$ is valued as$100 - 300$ MeV, and the running  constant in  soft region is 
$\alpha_s = \mathcal{O}(1)$\footnote{The effective intensity $Y(s)$ of hadron interactions is a dual quantity to the running  constant $\alpha_s$ in the soft region according to quark--hadron duality \cite{tandy}.}. The QCD-inspired hadron scattering models neglect the confinement   rather frequently.  

One can imagine that the color conducting medium is 
formed  instead of the color insulating one when the energy of  interacting hadrons
exceeds some threshold value and  associate appearance of the
reflective scattering mode with formation of this color conducting medium in the
intermediate state \cite{jpg}.  Of course, this is  not a unique interpretation of the reflective scattering mode, but due to its correspondence to the strong coupling  in hadron interactions, connection with    phenomena of confinement seems worthy of study i.e.
the reflective scattering mode can be interpreted as a consequence of confinement in QCD.

 \section*{Acnowledgement}
 We are grateful to Rami Oueslati for the interesting correspondence on the microscopic interpretations of  reflective scattering mode and to Vladimir Petrov for discussions of  eikonal approximation applicability.

\end{document}